\documentclass[%
 aip,
 jcp,%
 amsmath,amssymb,
 reprint,%
]{revtex4-1}

\usepackage[
top    = 1.8cm,
bottom = 1.5cm,
left   = 2cm,
right  = 2cm]{geometry}

\usepackage{graphicx}
\usepackage{dcolumn}
\usepackage{bm}

\begin{document}


\title{Local structure-mobility relationships of confined fluids reverse upon supercooling}

\author{Jonathan A. Bollinger}
\affiliation{McKetta Department of Chemical Engineering, University of Texas at Austin, Austin, Texas 78712, USA}

\author{Avni Jain}
\affiliation{McKetta Department of Chemical Engineering, University of Texas at Austin, Austin, Texas 78712, USA}

\author{James Carmer}
\affiliation{McKetta Department of Chemical Engineering, University of Texas at Austin, Austin, Texas 78712, USA}

\author{Thomas M. Truskett}
\email{truskett@che.utexas.edu}
\affiliation{McKetta Department of Chemical Engineering, University of Texas at Austin, Austin, Texas 78712, USA}

\date{\today}

\begin{abstract}

We examine the structural and dynamic properties of confined binary hard-sphere mixtures designed to mimic realizable colloidal thin films. Using computer simulations, governed by either Newtonian or overdamped Langevin dynamics, together with other techniques including a Fokker-Planck equation-based method, we measure the position-dependent and average diffusivities of particles along structurally isotropic and inhomogeneous dimensions of the fluids. At moderate packing fractions, local single-particle diffusivities normal to the direction of confinement are higher in regions of high total packing fraction; however, these trends are reversed as the film is supercooled at denser average packings. Auxiliary short-time measurements of particle displacements mirror data obtained for experimental supercooled colloidal systems. We find that average dynamics can be approximately predicted based on the distribution of available space for particle insertion across orders of magnitude in diffusivity regardless of the governing microscopic dynamics.

\end{abstract}

\pacs{Valid PACS appear here}
\maketitle



Confined fluids exhibit inhomogeneous structural and relaxation properties, which are general features of materials subjected to position-dependent external fields. Because confined fluids emerge in a diverse array of natural and technological contexts (e.g., water in biological media, polymer thin films, etc.), considerable attention has been directed at understanding how their static and dynamic properties relate to bulk fluid physics observed under similar conditions. As a result, the static properties of confined fluids, such as local one-body density \(\rho(z)\), are now well-understood in terms of physical intuition (e.g., emergence of particle layering near boundaries to relieve packing frustration~\cite{Kjellander1991,Israelachvili2011}) and can be predicted using microscopic approaches like density functional theory~\cite{Wu2007,Roth2010}. However, much less is understood about what controls the dynamics of inhomogeneous fluids, and only recently have efforts broadened to include developing theories~\cite{Krakoviack2005,Biroli2006,Archer2006,Lang2010,Olivares2013,Lang2012,Lang2014} and other tools~\cite{Hummer2007,Mittal2008,Nugent2007,Hinczewski2010,Carmer2014} for characterizing particle dynamics both on a spatially-averaged basis and as a function of position.


Given the difficulty of applying first principles to understand the dynamics of such systems, progress has been made by virtue of use pragmatic approaches, e.g., application and testing of semiempirical, quasi-universal scaling laws that relate transport coefficients of interest to static properties~\cite{Rosenfeld1977,Rosenfeld1999,Hoyt2000,Sharma06,Mittal2006s2,Mittal2006supercooled,GnanDyre2009,Krekelberg2009,PondJCP2011,Nayar13,Dyre2014}.
To wit, it has been shown that single-particle diffusivities, relaxation times, and viscosities along structurally-invariant (i.e., isotropic) dimensions of simple confined fluids can be predicted based on knowledge of how dynamic properties of the bulk fluid relate to static quantities including excess entropy \(s^{\text{ex}}\) (relative to the ideal gas) and fractional available space \(\exp\{c^{\text{(1)}}\}\) (or insertion probability \(p_0\)), which characterize short-range static correlations and particle packings, respectively~\cite{Mittal2006,Mittal2007,Mittal2007mixtures,Goel2008,Goel2009,Chopra2010confined,Borah12,Ma2013,Liu2013,Ingebrigtsen2013}.

In this spirit, one might expect that local particle mobility in an inhomogeneous fluid should similarly correlate with position-dependent static properties; in other words, the way particles navigate through the inhomogeneous environment might be encoded in the physics of motion observed in a bulk, homogeneous fluid. However, the validity of such a connection has yet to be carefully and systematically evaluated.
Despite providing other important insights, previous investigations directly measuring inhomogeneous dynamics have studied a variety of fluids governed by disparate interactions, external fields, and conditions, and they have also used different protocols to characterize the dynamics~\cite{Nugent2007,Mittal2008,Edmond2012,Olivares2013,Lang2014,Bollinger2014, BollingerJPCB2014}. As a result, even fundamental questions related to confined (and more generally inhomogeneous) fluids remain open: Do local and average correlations between particle mobility and structure universally reflect bulk behaviors? Do new structure-mobility relations emerge as inhomogeneous fluids are supercooled toward glass transitions? And does the choice of microscopic dynamics affect these qualitative trends?


\begin{figure}[b]
  \includegraphics[width=8.255cm]{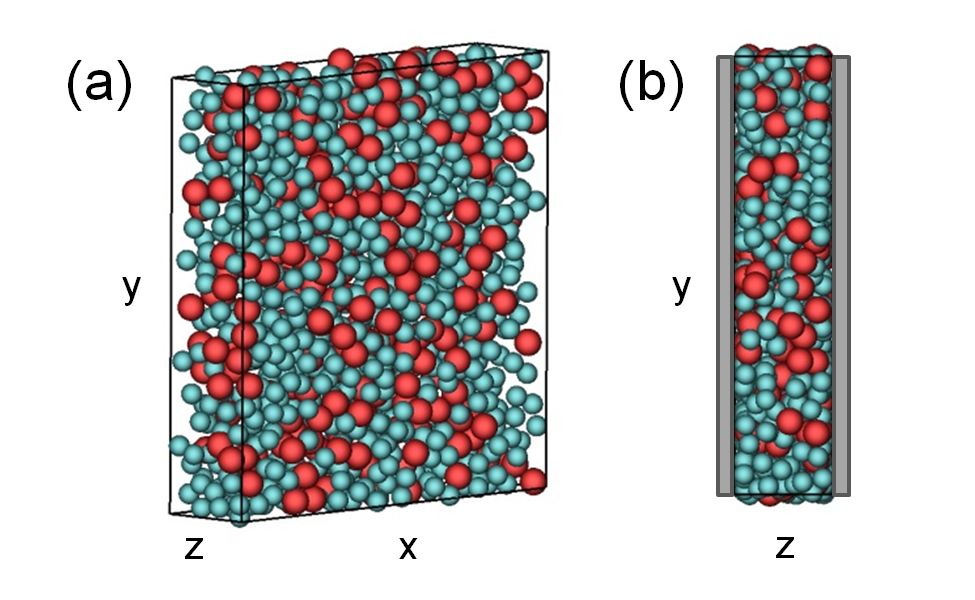}
  \caption{(color online). Illustration of confined binary mixture comprising small (blue) and large (red) particles.}
  \label{sch:Figure1}
\end{figure}


As a step toward addressing these questions, we examine computer simulations of bulk and confined binary mixtures of small (sm) and large (lg) hard spheres (HS) approximated by a steeply-repulsive Weeks-Chandler-Andersen (WCA) pair potential~\cite{ChandlerWeeksAndersenScience1983} between particles \(i\) and \(j\), adapted for multiple particle diameters: \(\varphi_{i,j}(r)=4\epsilon([\sigma_{\text{sm}}/(r+\Delta)]^{48}-[\sigma_{\text{sm}}/(r+\Delta)]^{24})+\epsilon\) for \(r\leq(2^{1/24}\sigma_{\text{sm}}-\Delta)\) and \(\varphi_{i,j}(r)=0\) for \(r>(2^{1/24}\sigma_{\text{sm}}-\Delta)\), where \(\epsilon\) is the characteristic energy scale; \(r\) is the interparticle separation; \(\sigma\) denotes particle diameter; and \(\Delta = \sigma_{\text{sm}} - (1/2)(\sigma_i+\sigma_j)\). The binary mixtures are composed of spheres with size ratio \(\sigma_{\text{lg}}/\sigma_{\text{sm}} = 1.3\), volume-proportional masses \(m_{\text{lg}}/m_{\text{sm}} = (\sigma_{\text{lg}}/\sigma_{\text{sm}})^3\), and composition defined by the fraction of small particles \(x_{\text{sm}} = 0.75\).  These parameters mimic colloidal mixtures investigated in recent experiments~\cite{Nugent2007,Edmond2012}. Below, we implicitly non-dimensionalize quantities via appropriate combinations of the characteristic lengthscale \(\sigma_{\text{sm}}\) and energy scale \(\epsilon=k_{\text{B}}T\), where \(k_{\text{B}}\) is Boltzmann's constant, and \(T\) is temperature.

Periodic boundary conditions are applied in all directions for the bulk systems, while for the confined systems (see Fig. 1), particles are situated in slit-pores of size \(H=5\) between two reflective walls placed at \(z = \pm H/2\), with periodic boundary conditions applied in the \(x\)- and \(y\)-directions. The wall-particle interactions are analogous to the hard-sphere-like interactions between particles, but defined such that the center of particle \(i\) can access \(-(H-\sigma_i)/2 \lesssim z \lesssim (H-\sigma_i)/2\). Spatially averaged packing fractions are given by \(\phi^{\text{avg}} = (\pi/6)\rho^{*}[x_{\text{sm}} + \sigma_{\text{lg}}^3 (1 - x_{\text{sm}})]\), where \(\rho^{*}=(N_{\text{sm}}+N_{\text{lg}})/V\) is the combined number density of both species and \(V\) is volume. Here, the \(\phi^{\text{avg}}\) values for the confined fluids are defined by the total (surface- rather than center-accessible) slit pore volume. We generate particle trajectories governed by either conventional molecular dynamics (MD) or Brownian dynamics (BD) (i.e., overdamped Langevin ignoring hydrodynamic interactions) using GROMACS 4.5.5~\cite{HessJCTC2008} with implementation details are provided in the Supplemental Material (SM)~\footnote{See supplemental material at [URL will be inserted by AIP] for detailed simulation protocols and auxiliary data.}.

To characterize particle motions, we calculate mean-squared displacements (MSDs) and diffusivities in the structurally isotropic \emph{and} inhomogeneous directions of the bulk and confined systems. Average diffusivities \(D^{\text{avg}}\) in the bulk systems and parallel to the walls in the confined systems characterize motions in isotropic directions, and are derived by fitting the long-time behavior of the MSD of all the particles to the Einstein relation \(\langle\Delta\mathbf{r}^{2}\rangle = 2dD\Delta t\). In the bulk (confined) case, \(\langle\Delta\mathbf{r}^{2}\rangle\) is the MSD in the \(x\)-, \(y\)-, and \(z\)-directions  (\(x\)- and \(y\)-directions) over lag-time \(\Delta t\) and dimensionality \(d=3\) (\(d=2\)).

Diffusivities in the inhomogeneous \(z\)-direction of the confined pores are position-dependent and cannot be calculated via the Einstein relation because particles are subjected to locally non-cancelling potentials of mean force~\cite{LiuBerne2004}. Particle displacements along the \(z\)-coordinate are instead accurately described~\cite{Mittal2008} by the 1D Fokker-Planck (FP) equation
\begin{equation}
\dfrac{\partial G}{\partial t} = \dfrac{\partial}{\partial z} \Bigg( D_{\text{z}}(z)e^{-F(z)}\dfrac{\partial}{\partial z}[e^{F(z)}G]\Bigg)
\end{equation}
where \(D_{\text{z}}(z)\) are position-dependent diffusivities. Here, \(G(z, t_0+\Delta t|z', t_0)\) is the Markovian propagator characterizing temporal single-particle displacements given the potential of mean force \(F(z)=-\ln\{\rho(z)\} + C\), where \(C\) is an arbitrary constant. To obtain \(D_{\text{z}}(z)\) from simulation data, we use a mean-first passage times (MFPT) method~\cite{WeissACP1966, Hinczewski2010, Netz2011} applied to the steady-state (i.e., \(\partial G/\partial t = 0\)) limit of the FP equation, which is known to provide equivalent information compared to alternative FP treatments~\cite{Mittal2008, Carmer2014}. Additional implementation details are discussed elsewhere~\cite{Bollinger2014,BollingerJPCB2014}.


\begin{figure}
  \includegraphics{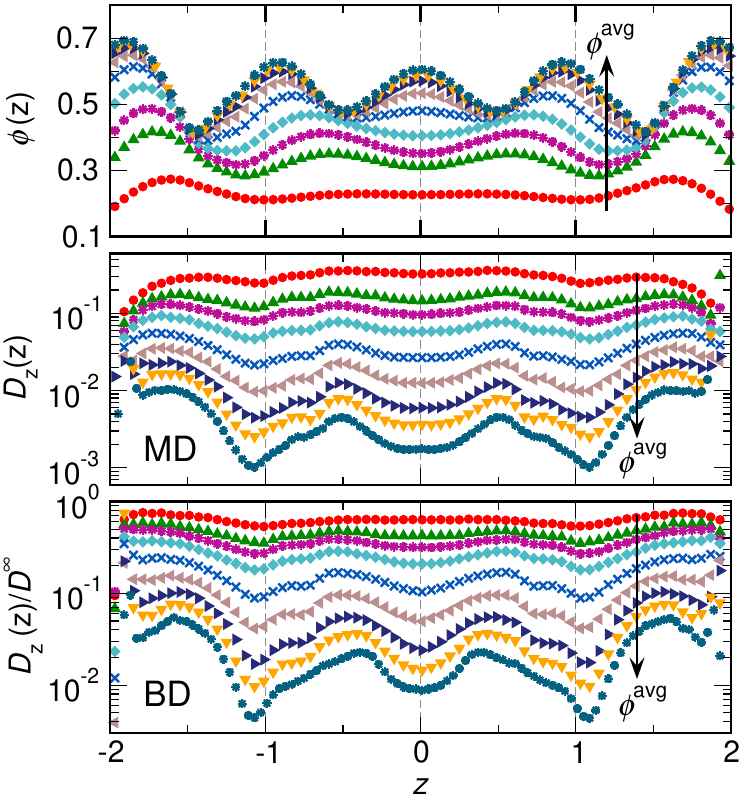}
  \caption{(color online). Local total packing fractions \(\phi(z)\) (top) and local diffusivities in the \(z\)-direction \(D_{\text{z}}(z)\) of small particles calculated from MD simulations (middle) and BD simulations (bottom) of pore size \(H=5\) and average total packing fractions \(\phi^{\text{avg}}\) = 0.20, 0.30, 0.35, 0.40, 0.45, 0.48, 0.50, 0.51, and 0.52. The BD profiles are normalized by the infinite dilution diffusivity \(D^{\infty}_{\sigma_{\text{sm}}}\). \(D_{\text{z}}(z)\) profiles for the large particles exhibit shapes in line with the small-particle profiles, as shown in the SM~\cite{Note1}.
  }
  \label{sch:Figure2}
\end{figure}

We begin our discussion by considering Fig. 2, where we compare local total packing fraction \(\phi(z)\) and local particle diffusivities in the \(z\)-direction \(D_{\text{z}}(z)\) for confined systems over a wide range of \(\phi^{\text{avg}}\) and governed by either Newtonian or Brownian microscopic dynamics. Here, we use \(\phi(z)\) because it more economically quantifies the local aggregate packing frustration compared to component density profiles \(\rho(z)\), where the latter are provided in the SM~\cite{Note1}. Remarkably, we find that while the packing structure in the confined pores undergoes an apparent shift from four to five dense particle layers upon increasing \(\phi^{\text{avg}}\), {\em the shapes of the \(D_{\text{z}}(z)\) profiles are qualitatively insensitive to this considerable structural rearrangement}. Thus, for \(\phi^{\text{avg}} \leq 0.40\), particles diffuse more quickly though densely-packed regions (except very close to the walls, where particles slow down due to impenetrability), but for \(\phi^{\text{avg}} \geq 0.45\), particles instead move more slowly through densely-packed regions.

This gradual reversal from positive to negative local correlations between packing fraction and mobility bridges observations based on previous measurements of particle dynamics in confined pores--measurements that seemingly pointed to inconsistent local trends, but where comparisons were also complicated by different protocols and dynamic regimes. Mittal et al.~\cite{Mittal2008} measured local FP-based diffusivities in Newtonian HS simulations at \emph{equilibrium} conditions (\(\phi^{\text{avg}} \leq 0.40\)) and observed positive correlations between local density \(\rho(z)\) (or \(\phi(z)\)) and \(D_{\text{z}}(z)\). In contrast, Nugent and co-workers~\cite{Nugent2007,Edmond2012} experimentally measured short-time MSDs along the \(z\)-coordinate as a function of position for \emph{supercooled} thin films of pseudo-HS colloids, results which pointed to negative correlations between local density and mobility. While the latter results more intuitively correlate to expectations based on bulk HS density trends, Mittal et. al. provide a plausible physical basis for the observed positive correlations. Specifically, they correctly note that higher-density regions in such inhomogeneous HS systems also exhibit the greater fraction of locally available space for inserting additional particle centers, i.e., more locally free volume, which might correlate with dynamics~\cite{Widom1963,Kjellander1991,Sastry1998}.

\begin{figure}
  \includegraphics{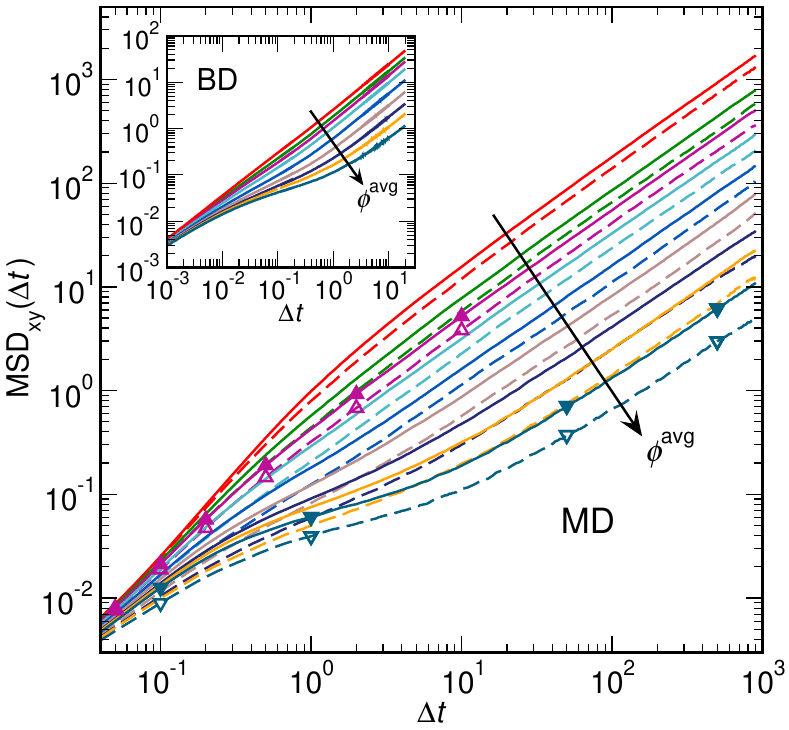}
  \caption{(color online). Mean-squared displacements (MSD) per particle in the \(x\)- and \(y\)-directions versus lag-times \(\Delta t\) for small (solid lines) and large (dashed lines) particles from MD (main) and BD (inset) simulations of pore size \(H=5\) and average total packing fractions \(\phi^{\text{avg}}\) = 0.20, 0.30, 0.35, 0.40, 0.45, 0.48, 0.50, 0.51, and 0.52. In (b), only small-particle curves are shown for clarity and lag-times have been normalized by \(D^{\infty}_{\sigma_{\text{sm}}} / \sigma_{\text{sm}}^2\). Symbols in (a) denote times corresponding to the profiles in Fig. 4.}
  \label{sch:Figure3}
\end{figure}

The results in Fig. 2 clearly demonstrate within a single framework for measuring dynamics that \emph{either} positive or negative correlations between density and diffusivity can be observed in these systems depending on whether equilibrium or supercooled conditions are being studied.  By considering Fig. 2 in conjunction with Fig. 3, where the latter shows particles MSDs parallel to the confining walls as a function of lag-time \(\Delta t\) and \(\phi^{\text{avg}}\), we observe that the reversal in local structure-mobility correlations approximately coincides with the emergence of plateaus in the MSDs at \(\phi^{\text{avg}} \gtrsim  0.45\), a signature of sub-diffusive ``particle caging'' characteristic of supercooling~\cite{WeeksWeitz2002,Nugent2007}.

The above results imply that {\em local} packing structure as measured by \(\phi(z)\) does not generally correlate in a nontrivial way to position-dependent diffusive mobility (this is also true of more ``microscopic'' local static quantities like \(p_{0}(z)\), as shown in the SM~\cite{Note1}). In turn, given that bulk HS fluids exhibit simple negative correlations between packing fraction and mobility, it is apparent that local static-dynamic correlations in confined-fluid systems cannot be na\"{i}vely extrapolated (or predicted) from the bulk physics, in agreement with findings for more idealized density-varying HS systems~\cite{Bollinger2014}. Interestingly, for the systems examined here, the choice of microscopic dynamics had no qualitative impact on the shapes of the \(D_{\text{z}}(z)\) profiles, though recent results~\cite{BollingerJPCB2014} indicate this is not generally true of inhomogeneous fluids.

\begin{figure}
  \includegraphics{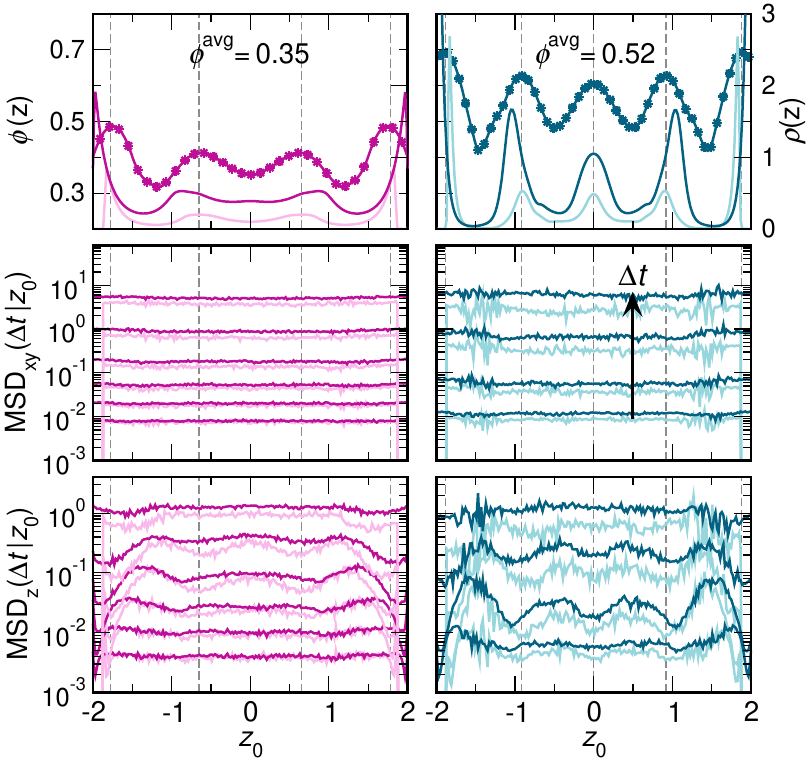}
  \caption{(color online). Local total packing fractions \(\phi(z)\) and individual component densities \(\rho(z)\) (top) and MSDs per particle in the \(x\)- and \(y\)-directions (middle) and \(z\)-direction (bottom) for various lag-times \(\Delta t\) plotted as a function of particle position \(z_0\) at \(\Delta t = 0\). Apart from \(\phi(z)\) profiles (line-symbols), results for small and large particles are plotted with darker and lighter curves, respectively. Results are calculated from MD simulations for pore size \(H=5\), where left panels show results for \(\phi^{\text{avg}}\) = 0.35 and \(\Delta t\) = 0.05, 0.1, 0.2, 0.5, 2.0, and 10.0 and right panels for \(\phi^{\text{avg}}\) = 0.52 and \(\Delta t\) = 0.1, 1.0, 50.0, and 500.0. Lag-times are also plotted in Fig. 3 as symbols.}
  \label{sch:Figure4}
\end{figure}

We next provide data reinforcing the idea that the opposing correlations in Fig. 2 between packing fraction and diffusivity at equilibrium versus supercooled conditions may also emerge in real colloidal thin films: in Fig. 4, we compare position- and time-dependent particles MSDs of the confined films against \(\phi(z)\) and component \(\rho(z)\), results that mirror experimental measurements by Nugent and co-workers (see, e.g., Figs. 7-8 from~\cite{Edmond2012}) for supercooled thin films of pseudo-HS. In particular, we show results calculated from MD simulations for \(\phi^{\text{avg}}=0.35\) and 0.52, which correspond to equilibrium and supercooled conditions, respectively.

Mirroring the experimental findings, for the supercooled conditions in Fig. 4, MSDs in the \(xy\)-plane are insensitive with respect to originating position \(z_0\) (i.e., position at lag-time \(\Delta t = 0\)) for all \(\Delta t\), while MSDs in the inhomogeneous \(z\)-direction are negatively correlated with respect to \(\phi(z)\) (\(\rho(z)\)) for sufficiently short \(\Delta t\). At longer \(\Delta t\), the MSD dependence on \(z_0\) disappears as particles are no longer generally situated near \(z_0\). At \(\phi^{\text{avg}}=0.35\), MSDs in the \(xy\)-plane also do not vary with \(z_0\), but MSDs in the \(z\)-direction are instead positively correlated with respect to \(\phi(z)\) (\(\rho(z)\)) for \(\Delta t \leq 0.2\), though the correlations appear to reverse at longer \(\Delta t \geq 0.50\) before washing out at long lag-times. To our knowledge, no analogous data for real non-supercooled thin films for has yet been published.

Surprisingly, the MSDs in the \(z\)-direction at many \(\Delta t\) approximately reflect the \(D_{\text{z}}(z)\) profiles in Fig. 2 even though the quantified motions are by necessity sub-diffusive and accrued when a particle is no longer at \(z_0\). In turn, only profiles at the \emph{very shortest times} provide information about motions precisely at \(z_0\), but these are also furthest from the diffusive regime. Nonetheless, given that the results for the supercooled system in Fig. 3 are consistent with the available experimental data, it is plausible that the FP-derived results for diffusive motions--including the positive correlations between \(\phi(z)\) and \(D_{\text{z}}(z)\) and their reversal at high \(\phi^{\text{avg}}\)--can be observed in real confined colloids.

\begin{figure}
  \includegraphics{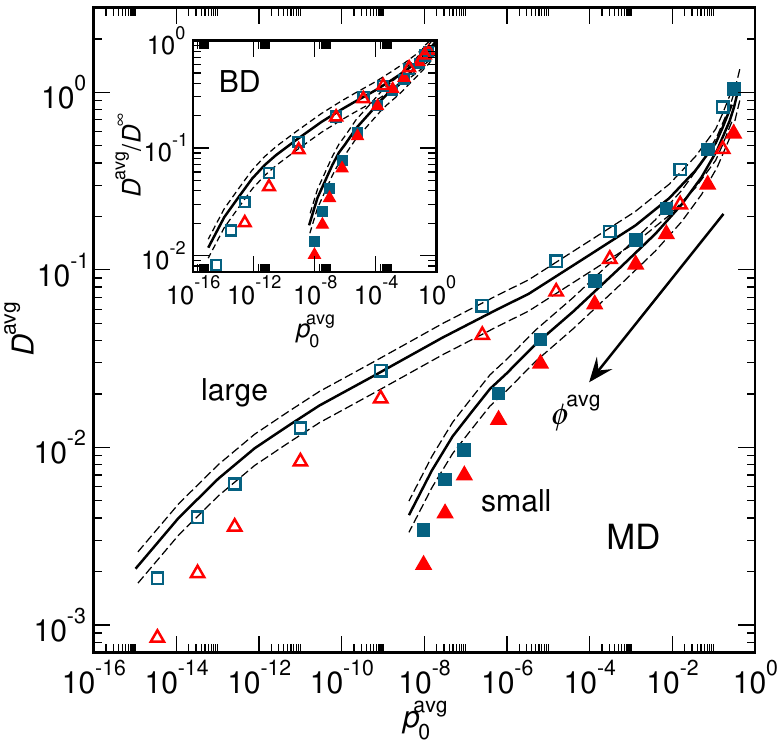}
  \caption{(color online). Average component diffusivities \(D^{\text{avg}}\) in the \(xy\)-plane (squares) and \(z\)-direction (triangles) versus average component insertion probabilities \(p_{\text{0}}^{\text{avg}}\) for small and large particles (filled and unfilled symbols, respectively) calculated from MD simulations (main) and BD simulations (inset) of pore size \(H=5\) and average total packing fractions \(\phi^{\text{avg}}\) = 0.20, 0.30, 0.35, 0.40, 0.45, 0.48, 0.50, 0.51, and 0.52. Average diffusivities for bulk mixtures shown as solid black lines with \(\pm 20 \%\) bounds shown as dashed lines.}
  \label{sch:Figure5}
\end{figure}

Given that the previous results undermine any notion of a universal connection between local structure and local mobility, it is natural to wonder whether \emph{average} diffusivities in directions parallel and perpendicular to the confining walls reflect bulk fluid physics and can be predicted based on average static properties. In Fig. 5, we address this by comparing average diffusivities \(D^{\text{avg}}_{\text{xy}}\) and \(D^{\text{avg}}_{\text{z}}(z) = \int_0^{H/2}D_{\text{z}}(z)\rho(z)\mathrm{d}z\text{ }/\int_0^{H/2}\rho(z)\mathrm{d}z\) for the small and large particles from the confined pores against curves for bulk mixtures. Here, we plot these dynamic quantities against component-specific average insertion probabilities \(p_{0}^{\text{avg}}\) (or available volumes for insertion), which have been shown to provide the most quantitatively robust connection between bulk diffusivity and \(D_{\text{xy}}\) in confined slit pores of HS governed by Newtonian dynamics~\cite{Goel2009}. Details of \(p_{0}^{\text{avg}}\) calculations are provided in the SM~\cite{Note1}.


As is evident in Fig. 5, the average diffusivities \(D_{\text{xy}}\) and \(D^{\text{avg}}_{\text{z}}(z)\) of the confined fluids approximately collapse onto the relevant bulk curves over many orders of magnitude in \(p_{0}^{\text{avg}}\) for systems governed by either Newtonian and Brownian dynamics. Notably, even at high \(\phi^{\text{avg}}\) associated with supercooling, \(D^{\text{avg}}_{\text{z}}(z)\) values only differ from the bulk by factors of 2-3 based on component \(p_{0}^{\text{avg}}\); if one instead plots diffusivities against a less ``microscopic'' static property, e.g., component \(\rho^{\text{avg}}\), confined and bulk diffusivities differ by up to an order of magnitude. Overall, the data support the idea that, despite the difficulty of rationalizing position-dependent diffusivity behaviors based on bulk physics, the average dynamics of inhomogeneous fluids are nonetheless strongly encoded with bulk correlations between mobility and available space.


In closing, by characterizing the particle dynamics of highly confined binary HS mixtures in both inhomogeneous and isotropic dimensions, we find that diffusive mobility is not universally predicated upon packing structure according to bulk HS behaviors, as exemplified by the reversal from positive to negative correlations between local total packing fraction \(\phi(z)\) and single-particle diffusivity \(D^{\text{avg}}_{\text{z}}(z)\) coinciding with the onset of supercooling. In contrast, average diffusive mobility is strongly encoded by the bulk physics, and can be approximately predicted via knowledge of the distribution of available space. For the confined fluids studied here, results are insensitive to whether Newtonian or Brownian (i.e., overdamped Langevin) microscopic dynamics govern particle trajectories, though it is an open question as to whether similar classes of behavior will emerge in real colloidal thin films treated within the FP formalism. More speculatively, the shapes of the \(D^{\text{avg}}_{\text{z}}(z)\) profiles (and their qualitative insensitivity to \(\phi^{\text{avg}}\)) suggest that there may simply be a ``universal'' oscillatory signature of local diffusivity that emerges for non-continuum fluids proximal to confining potentials regardless of microscopic dynamics or the specific nature of any emergent structural inhomogeneity. We are presently investigating this possibility.

\section*{ACKNOWLEDGMENTS}

\noindent We kindly thank Dr. Vincent Shen for providing thermodynamic data for the bulk and confined systems. This work was supported by the Robert A. Welch Foundation (F-1696), and the National Science Foundation (CBET-1403768). We also acknowledge the Texas Advanced Computing Center (TACC) at The University of Texas at Austin for providing HPC resources for this study.

\section*{APPENDIX A: Simulation Protocols}

To generate particle trajectories governed by either molecular dynamics (MD) or Brownian dynamics (BD), we simulate systems of \(N_{\text{sm}}+N_{\text{lg}}=2400\) particles using GROMACS 4.5.5~\cite{HessJCTC2008}. MD trajectories are generated by integrating the Newtonian equations of motion with a time step of 0.001 while fixing temperature with a Nose-Hoover thermostat. BD trajectories are generated via the overdamped Langevin equation (ignoring hydrodynamic interactions), where the position \(\mathbf{r}_i\) of particle \(i\) is propogated with a time-step of 0.01 according to~\cite{Ermak1978, AllenTildesley1989}: \(\mathbf{r}_{i}(t+\Delta t) = \mathbf{r}_{i}(t) + D^{\infty}_{\sigma_{i}} \Delta t \mathbf{F}_{i}(\mathbf{r}_{i}(t)) + \boldsymbol{\xi}_{i}(t)\). Here, \(D^{\infty}_{\sigma_{i}}\) is the infinite dilution diffusivity, \(\mathbf{F}_{i}(t)\) is the net force due to interparticle and wall interactions, and \(\boldsymbol{\xi}_{i}(t)\) is the stochastic contribution. We set \(D^{\infty}_{\sigma_{\text{sm}}} = 0.001\) and \(D^{\infty}_{\sigma_{\text{lg}}}/D^{\infty}_{\sigma_{\text{sm}}} = \sigma_{\text{sm}}/\sigma_{\text{lg}}\), and in each direction, \(\xi_{i}(t) = r^\text{G}(t)\sqrt{2 D^{\infty}_{\sigma_{i}} \Delta t} \), where \(r^\text{G}(t)\) is a Gaussian noise with \(\langle r^\text{G}(t) \rangle = 0\) and variance \(\sigma^2 = 1\).

To generate bulk and confined packings at high \(\phi^{\text{avg}}\), we initialize systems at \(\phi^{\text{avg}}<0.30\) and compress them to the desired packing fractions via the method of Lubachevsky and Stillinger~\cite{LubachevskyStillinger1990}, in which particle diameters are grown linearly with time according to the dimensionless growth rate \(\Gamma\). We execute compressions via MD simulations with effective \(\Gamma < 1{\text{x}}10^{-6}\), which allows us to avoid generating partially jammed (i.e., non-equilibrated) structures for all presented \(\phi^{\text{avg}}\). Further equilibration and production runs (MD and BD) are then initialized with the final structures.

\section*{APPENDIX B: Density profiles across confined pores}

\begin{figure}
  \includegraphics{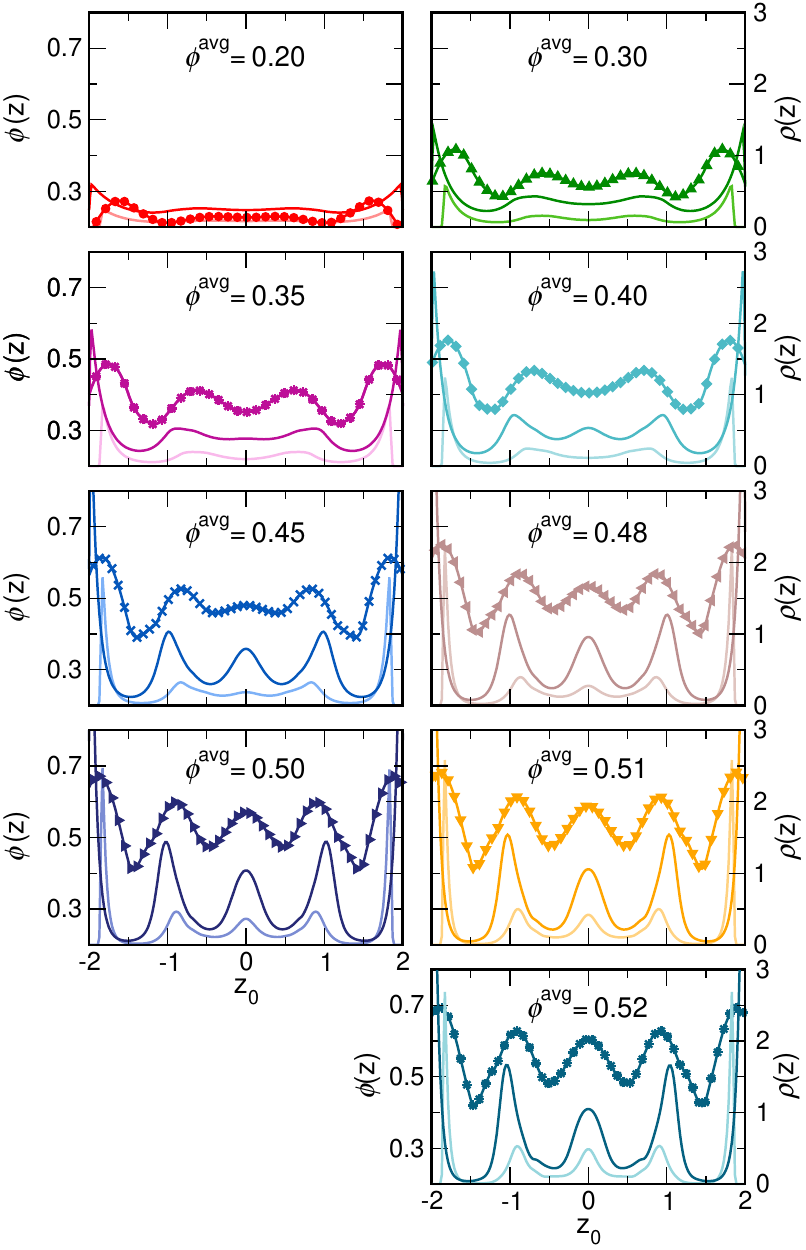}
  \caption{(color online). Local total packing fractions \(\phi(z)\) (line-symbols) and individual component densities \(\rho(z)\) (lines) from simulations of pore size \(H=5\) at various average total packing fractions \(\phi^{\text{avg}}\), where results for small and large particles are plotted with darker and lighter curves, respectively.}
  \label{sch:FigureSM1}
\end{figure}

In the main text, we largely plot local total packing fractions \(\phi(z)\) to characterize fluid structure. However, a few noteworthy aspects concerning fluid structure are apparent by simultaneously considering individual component density profiles \(\rho(z)\), as shown here in Fig. 1. First, for select conditions (e.g., \(\phi^{\text{avg}} = 0.40\)), a given individual component \(\rho(z)\) profile can be \emph{out of phase} with \(\phi(z)\). This does not happen frequently, and given the characteristics of the mixture (e.g., composition \(x_{\text{sm}} = 0.75\), size ratio \(\sigma_{\text{lg}}/\sigma_{\text{sm}} = 1.3\)) studied here, this behavior is only observed for small particle profiles. This underlines the importance of characterizing the spatial distribution of both particle species in order to gain a complete picture of packing structure; otherwise, one may come to \emph{qualitatively} incorrect conclusions about which regions of a fluid are densely packed. However, for the size ratio studied here, qualitative variations in \(\phi(z)\) can mostly be derived from knowledge only of the \emph{large} particle \(\rho(z)\) profiles (as would likely be possible for all \(x_{\text{sm}} \leq 0.75\)).

\begin{figure}
  \includegraphics{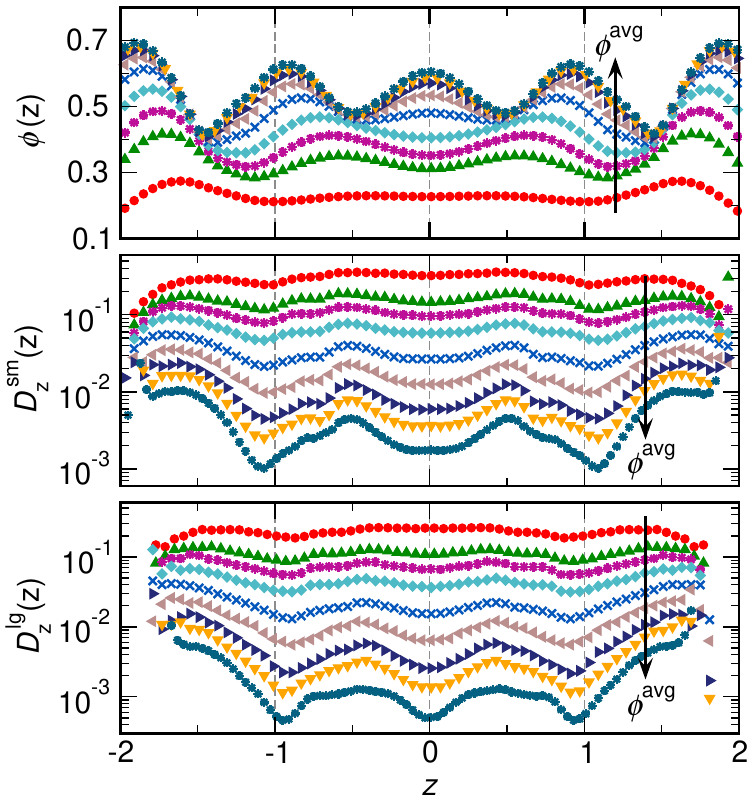}
  \caption{(color online). Local total packing fractions \(\phi(z)\) (top) and local diffusivities in the \(z\)-direction \(D_{\text{z}}(z)\) of small (middle) and large (bottom) particles calculated from MD simulations of pore size \(H=5\) and average total packing fractions \(\phi^{\text{avg}}\) = 0.20, 0.30, 0.35, 0.40, 0.45, 0.48, 0.50, 0.51, and 0.52.}
  \label{sch:FigureSM2}
\end{figure}

As discussed in relation to Figs. 2 and 5 in the main text, by comparing these \(\rho(z)\) profiles against local particle diffusivities \(D^{\text{avg}}_{\text{z}}(z)\), it is evident that there is no \(\phi^{\text{avg}}\)-independent local correlation between component density and diffusive mobility. Likewise, the local available space for particle insertion, as quantified by insertion probability \(p_{0}(z)\) in these systems, does not universally correlate with mobility because (see Section IV) \(p_{0}(z) = \rho(z)/\xi\), where \(\xi\) is the spatially-invariant component activity~\cite{Widom1963}. Of course, this implies that other position-dependent measures of structural correlations that are positively correlated with \(p_{0}(z)\), such as the local two-body excess entropy \(s^\text{(2)}(z)\), would likewise exhibit no consistent correlation with diffusive mobility~\cite{Bollinger2014}. Taken altogether, the results in the supplemental Fig. 1 underline that idea that local structure does not control, nor can be used to predict, local diffusive mobility in any straightforward way.

\section*{APPENDIX C: Component local diffusivities}

In supplemental Fig. 2, we show local particle diffusivities \(D^{\text{avg}}_{\text{z}}(z)\) for both small and large particles governed by MD at the various \(\phi^{\text{avg}}\) conditions studied. As noted in the main text, it is evident that the \(D^{\text{avg}}_{\text{z}}(z)\) profiles for the two types of particles have qualitatively similar shapes and exhibit the same insenstivity to the structural rearrangment that occurs upon the onset of supercooling at high \(\phi^{\text{avg}}\). Note that there is slight drift (i.e., overestimaton of \(D^{\text{avg}}_{\text{z}}(z)\)) near the edges of the large-particle profiles at higher \(\phi^{\text{avg}}\) due to the relatively meager amount of particle trajectory data obtained for this species (\(x_{\text{lg}} = 0.25\)). \(D^{\text{avg}}_{\text{z}}(z)\) profiles for small and large particles governed by overdamped Langevin dynamics are also qualitatively similar.

\section*{APPENDIX D: Calculating average insertion probability}

To calculate \(p_{0}^{\text{avg}}\) for each component, we note that if there is no external field at position \(z\) (i.e., \(\varphi^{\text{ext}}(z) = 0\)), the \emph{local} insertion probability~\cite{HansenMcDonald2006} for bulk or inhomogeneous HS is a ratio~\cite{Widom1963} \(p_{0}(z) = \rho(z)/\xi\) of the local component density \(\rho(z)\) and the spatially-invariant component activity \(\xi = \exp(\beta\mu)/\lambda^3\), where the latter is defined by the component chemical potential \(\mu\) and the de Broglie wavelength \(\lambda\). Given that we have component \(\rho(z)\) profiles measured from the MD and BD simulations, all that is required to obtain \(p_{0}(z)\) profiles are activities \(\xi\) for the bulk and confined mixtures at the various \(\phi^{\text{avg}}\) values. These \(\xi\) values are obtained via grand canonical transition matrix Monte Carlo (GC-TMMC) simulations~\cite{Shen2005}, with implementation details presented elsewhere~\cite{Goel2009}. It is then straightforward to calculate \(p_{0}^{\text{avg}} = H^{-1} \int_0^{H/2}p_{0}(z)\mathrm{d}z\), which for bulk mixtures is simply \(p^{\text{avg}}_{0} = \rho/\xi\).


\begin{thebibliography}{54}%
\makeatletter
\providecommand \@ifxundefined [1]{%
 \@ifx{#1\undefined}
}%
\providecommand \@ifnum [1]{%
 \ifnum #1\expandafter \@firstoftwo
 \else \expandafter \@secondoftwo
 \fi
}%
\providecommand \@ifx [1]{%
 \ifx #1\expandafter \@firstoftwo
 \else \expandafter \@secondoftwo
 \fi
}%
\providecommand \natexlab [1]{#1}%
\providecommand \enquote  [1]{``#1''}%
\providecommand \bibnamefont  [1]{#1}%
\providecommand \bibfnamefont [1]{#1}%
\providecommand \citenamefont [1]{#1}%
\providecommand \href@noop [0]{\@secondoftwo}%
\providecommand \href [0]{\begingroup \@sanitize@url \@href}%
\providecommand \@href[1]{\@@startlink{#1}\@@href}%
\providecommand \@@href[1]{\endgroup#1\@@endlink}%
\providecommand \@sanitize@url [0]{\catcode `\\12\catcode `\$12\catcode
  `\&12\catcode `\#12\catcode `\^12\catcode `\_12\catcode `\%12\relax}%
\providecommand \@@startlink[1]{}%
\providecommand \@@endlink[0]{}%
\providecommand \url  [0]{\begingroup\@sanitize@url \@url }%
\providecommand \@url [1]{\endgroup\@href {#1}{\urlprefix }}%
\providecommand \urlprefix  [0]{URL }%
\providecommand \Eprint [0]{\href }%
\providecommand \doibase [0]{http://dx.doi.org/}%
\providecommand \selectlanguage [0]{\@gobble}%
\providecommand \bibinfo  [0]{\@secondoftwo}%
\providecommand \bibfield  [0]{\@secondoftwo}%
\providecommand \translation [1]{[#1]}%
\providecommand \BibitemOpen [0]{}%
\providecommand \bibitemStop [0]{}%
\providecommand \bibitemNoStop [0]{.\EOS\space}%
\providecommand \EOS [0]{\spacefactor3000\relax}%
\providecommand \BibitemShut  [1]{\csname bibitem#1\endcsname}%
\let\auto@bib@innerbib\@empty

\bibitem [{\citenamefont {Kjellander}\ and\ \citenamefont
  {Sarman}(1991)}]{Kjellander1991}%
  \BibitemOpen
  \bibfield  {author} {\bibinfo {author} {\bibfnamefont {R.}~\bibnamefont
  {Kjellander}}\ and\ \bibinfo {author} {\bibfnamefont {S.}~\bibnamefont
  {Sarman}},\ }\href {\doibase 10.1039/FT9918701869} {\bibfield  {journal}
  {\bibinfo  {journal} {J. Chem. Soc.{,} Faraday Trans.}\ }\textbf {\bibinfo
  {volume} {87}},\ \bibinfo {pages} {1869} (\bibinfo {year}
  {1991})}\BibitemShut {NoStop}%
\bibitem [{\citenamefont {Israelachvili}(2011)}]{Israelachvili2011}%
  \BibitemOpen
  \bibfield  {author} {\bibinfo {author} {\bibfnamefont {J.~N.}\ \bibnamefont
  {Israelachvili}},\ }\href@noop {} {\emph {\bibinfo {title} {Intermolecular
  and Surface Forces}}}\ (\bibinfo  {publisher} {Academic Press},\ \bibinfo
  {address} {New York, NY, USA},\ \bibinfo {year} {2011})\BibitemShut {NoStop}%
\bibitem [{\citenamefont {Wu}\ and\ \citenamefont {Li}(2007)}]{Wu2007}%
  \BibitemOpen
  \bibfield  {author} {\bibinfo {author} {\bibfnamefont {J.}~\bibnamefont
  {Wu}}\ and\ \bibinfo {author} {\bibfnamefont {Z.}~\bibnamefont {Li}},\ }\href
  {\doibase 10.1146/annurev.physchem.58.032806.104650} {\bibfield  {journal}
  {\bibinfo  {journal} {Annu. Rev. Phys. Chem.}\ }\textbf {\bibinfo {volume}
  {58}},\ \bibinfo {pages} {85} (\bibinfo {year} {2007})}\BibitemShut {NoStop}%
\bibitem [{\citenamefont {Roth}(2010)}]{Roth2010}%
  \BibitemOpen
  \bibfield  {author} {\bibinfo {author} {\bibfnamefont {R.}~\bibnamefont
  {Roth}},\ }\href {http://stacks.iop.org/0953-8984/22/i=6/a=063102} {\bibfield
   {journal} {\bibinfo  {journal} {J. Phys.: Condens. Matter}\ }\textbf
  {\bibinfo {volume} {22}},\ \bibinfo {pages} {063102} (\bibinfo {year}
  {2010})}\BibitemShut {NoStop}%
\bibitem [{\citenamefont {Krakoviack}(2005)}]{Krakoviack2005}%
  \BibitemOpen
  \bibfield  {author} {\bibinfo {author} {\bibfnamefont {V.}~\bibnamefont
  {Krakoviack}},\ }\href {http://stacks.iop.org/0953-8984/17/i=45/a=049}
  {\bibfield  {journal} {\bibinfo  {journal} {J. Phys.: Condens. Matter}\
  }\textbf {\bibinfo {volume} {17}},\ \bibinfo {pages} {S3565} (\bibinfo {year}
  {2005})}\BibitemShut {NoStop}%
\bibitem [{\citenamefont {Biroli}\ \emph {et~al.}(2006)\citenamefont {Biroli},
  \citenamefont {Bouchaud}, \citenamefont {Miyazaki},\ and\ \citenamefont
  {Reichman}}]{Biroli2006}%
  \BibitemOpen
  \bibfield  {author} {\bibinfo {author} {\bibfnamefont {G.}~\bibnamefont
  {Biroli}}, \bibinfo {author} {\bibfnamefont {J.-P.}\ \bibnamefont
  {Bouchaud}}, \bibinfo {author} {\bibfnamefont {K.}~\bibnamefont {Miyazaki}},
  \ and\ \bibinfo {author} {\bibfnamefont {D.~R.}\ \bibnamefont {Reichman}},\
  }\href {\doibase 10.1103/PhysRevLett.97.195701} {\bibfield  {journal}
  {\bibinfo  {journal} {Phys. Rev. Lett.}\ }\textbf {\bibinfo {volume} {97}},\
  \bibinfo {pages} {195701} (\bibinfo {year} {2006})}\BibitemShut {NoStop}%
\bibitem [{\citenamefont {Archer}(2006)}]{Archer2006}%
  \BibitemOpen
  \bibfield  {author} {\bibinfo {author} {\bibfnamefont {A.~J.}\ \bibnamefont
  {Archer}},\ }\href {http://stacks.iop.org/0953-8984/18/i=24/a=004} {\bibfield
   {journal} {\bibinfo  {journal} {J. Phys.: Condens. Matter}\ }\textbf
  {\bibinfo {volume} {18}},\ \bibinfo {pages} {5617} (\bibinfo {year}
  {2006})}\BibitemShut {NoStop}%
\bibitem [{\citenamefont {Lang}\ \emph {et~al.}(2010)\citenamefont {Lang},
  \citenamefont {Bo\ifmmode~\mbox{\c{t}}\else \c{t}\fi{}an}, \citenamefont
  {Oettel}, \citenamefont {Hajnal}, \citenamefont {Franosch},\ and\
  \citenamefont {Schilling}}]{Lang2010}%
  \BibitemOpen
  \bibfield  {author} {\bibinfo {author} {\bibfnamefont {S.}~\bibnamefont
  {Lang}}, \bibinfo {author} {\bibfnamefont {V.}~\bibnamefont
  {Bo\ifmmode~\mbox{\c{t}}\else \c{t}\fi{}an}}, \bibinfo {author}
  {\bibfnamefont {M.}~\bibnamefont {Oettel}}, \bibinfo {author} {\bibfnamefont
  {D.}~\bibnamefont {Hajnal}}, \bibinfo {author} {\bibfnamefont
  {T.}~\bibnamefont {Franosch}}, \ and\ \bibinfo {author} {\bibfnamefont
  {R.}~\bibnamefont {Schilling}},\ }\href {\doibase
  10.1103/PhysRevLett.105.125701} {\bibfield  {journal} {\bibinfo  {journal}
  {Phys. Rev. Lett.}\ }\textbf {\bibinfo {volume} {105}},\ \bibinfo {pages}
  {125701} (\bibinfo {year} {2010})}\BibitemShut {NoStop}%
\bibitem [{\citenamefont {Olivares-Rivas}, \citenamefont {Colmenares},\ and\
  \citenamefont {L\'opez}(2013)}]{Olivares2013}%
  \BibitemOpen
  \bibfield  {author} {\bibinfo {author} {\bibfnamefont {W.}~\bibnamefont
  {Olivares-Rivas}}, \bibinfo {author} {\bibfnamefont {P.~J.}\ \bibnamefont
  {Colmenares}}, \ and\ \bibinfo {author} {\bibfnamefont {F.}~\bibnamefont
  {L\'opez}},\ }\href {\doibase http://dx.doi.org/10.1063/1.4818533} {\bibfield
   {journal} {\bibinfo  {journal} {J. Chem. Phys.}\ }\textbf {\bibinfo {volume}
  {139}},\ \bibinfo {pages} {074103} (\bibinfo {year} {2013})}\BibitemShut
  {NoStop}%
\bibitem [{\citenamefont {Lang}\ \emph {et~al.}(2012)\citenamefont {Lang},
  \citenamefont {Schilling}, \citenamefont {Krakoviack},\ and\ \citenamefont
  {Franosch}}]{Lang2012}%
  \BibitemOpen
  \bibfield  {author} {\bibinfo {author} {\bibfnamefont {S.}~\bibnamefont
  {Lang}}, \bibinfo {author} {\bibfnamefont {R.}~\bibnamefont {Schilling}},
  \bibinfo {author} {\bibfnamefont {V.}~\bibnamefont {Krakoviack}}, \ and\
  \bibinfo {author} {\bibfnamefont {T.}~\bibnamefont {Franosch}},\ }\href
  {\doibase 10.1103/PhysRevE.86.021502} {\bibfield  {journal} {\bibinfo
  {journal} {Phys. Rev. E}\ }\textbf {\bibinfo {volume} {86}},\ \bibinfo
  {pages} {021502} (\bibinfo {year} {2012})}\BibitemShut {NoStop}%
\bibitem [{\citenamefont {Lang}\ and\ \citenamefont
  {Franosch}(2014)}]{Lang2014}%
  \BibitemOpen
  \bibfield  {author} {\bibinfo {author} {\bibfnamefont {S.}~\bibnamefont
  {Lang}}\ and\ \bibinfo {author} {\bibfnamefont {T.}~\bibnamefont
  {Franosch}},\ }\href {\doibase 10.1103/PhysRevE.89.062122} {\bibfield
  {journal} {\bibinfo  {journal} {Phys. Rev. E}\ }\textbf {\bibinfo {volume}
  {89}},\ \bibinfo {pages} {062122} (\bibinfo {year} {2014})}\BibitemShut
  {NoStop}%
\bibitem [{\citenamefont {Hummer}(2005)}]{Hummer2007}%
  \BibitemOpen
  \bibfield  {author} {\bibinfo {author} {\bibfnamefont {G.}~\bibnamefont
  {Hummer}},\ }\href {http://stacks.iop.org/1367-2630/7/i=1/a=034} {\bibfield
  {journal} {\bibinfo  {journal} {New J. Phys.}\ }\textbf {\bibinfo {volume}
  {7}},\ \bibinfo {pages} {34} (\bibinfo {year} {2005})}\BibitemShut {NoStop}%
\bibitem [{\citenamefont {Mittal}\ \emph {et~al.}(2008)\citenamefont {Mittal},
  \citenamefont {Truskett}, \citenamefont {Errington},\ and\ \citenamefont
  {Hummer}}]{Mittal2008}%
  \BibitemOpen
  \bibfield  {author} {\bibinfo {author} {\bibfnamefont {J.}~\bibnamefont
  {Mittal}}, \bibinfo {author} {\bibfnamefont {T.~M.}\ \bibnamefont
  {Truskett}}, \bibinfo {author} {\bibfnamefont {J.~R.}\ \bibnamefont
  {Errington}}, \ and\ \bibinfo {author} {\bibfnamefont {G.}~\bibnamefont
  {Hummer}},\ }\href {\doibase 10.1103/PhysRevLett.100.145901} {\bibfield
  {journal} {\bibinfo  {journal} {Phys. Rev. Lett.}\ }\textbf {\bibinfo
  {volume} {100}},\ \bibinfo {pages} {145901} (\bibinfo {year}
  {2008})}\BibitemShut {NoStop}%
\bibitem [{\citenamefont {Nugent}\ \emph {et~al.}(2007)\citenamefont {Nugent},
  \citenamefont {Edmond}, \citenamefont {Patel},\ and\ \citenamefont
  {Weeks}}]{Nugent2007}%
  \BibitemOpen
  \bibfield  {author} {\bibinfo {author} {\bibfnamefont {C.~R.}\ \bibnamefont
  {Nugent}}, \bibinfo {author} {\bibfnamefont {K.~V.}\ \bibnamefont {Edmond}},
  \bibinfo {author} {\bibfnamefont {H.~N.}\ \bibnamefont {Patel}}, \ and\
  \bibinfo {author} {\bibfnamefont {E.~R.}\ \bibnamefont {Weeks}},\ }\href
  {\doibase 10.1103/PhysRevLett.99.025702} {\bibfield  {journal} {\bibinfo
  {journal} {Phys. Rev. Lett.}\ }\textbf {\bibinfo {volume} {99}},\ \bibinfo
  {pages} {025702} (\bibinfo {year} {2007})}\BibitemShut {NoStop}%
\bibitem [{\citenamefont {Hinczewski}\ \emph {et~al.}(2010)\citenamefont
  {Hinczewski}, \citenamefont {von Hansen}, \citenamefont {Dzubiella},\ and\
  \citenamefont {Netz}}]{Hinczewski2010}%
  \BibitemOpen
  \bibfield  {author} {\bibinfo {author} {\bibfnamefont {M.}~\bibnamefont
  {Hinczewski}}, \bibinfo {author} {\bibfnamefont {Y.}~\bibnamefont {von
  Hansen}}, \bibinfo {author} {\bibfnamefont {J.}~\bibnamefont {Dzubiella}}, \
  and\ \bibinfo {author} {\bibfnamefont {R.~R.}\ \bibnamefont {Netz}},\ }\href
  {\doibase http://dx.doi.org/10.1063/1.3442716} {\bibfield  {journal}
  {\bibinfo  {journal} {J. Chem. Phys.}\ }\textbf {\bibinfo {volume} {132}},\
  \bibinfo {pages} {245103} (\bibinfo {year} {2010})}\BibitemShut {NoStop}%
\bibitem [{\citenamefont {Carmer}, \citenamefont {van Swol},\ and\
  \citenamefont {Truskett}(2014)}]{Carmer2014}%
  \BibitemOpen
  \bibfield  {author} {\bibinfo {author} {\bibfnamefont {J.}~\bibnamefont
  {Carmer}}, \bibinfo {author} {\bibfnamefont {F.}~\bibnamefont {van Swol}}, \
  and\ \bibinfo {author} {\bibfnamefont {T.~M.}\ \bibnamefont {Truskett}},\
  }\href {\doibase http://dx.doi.org/10.1063/1.4890969} {\bibfield  {journal}
  {\bibinfo  {journal} {J. Chem. Phys.}\ }\textbf {\bibinfo {volume} {141}},\
  \bibinfo {pages} {046101} (\bibinfo {year} {2014})}\BibitemShut {NoStop}%
\bibitem [{\citenamefont {Rosenfeld}(1977)}]{Rosenfeld1977}%
  \BibitemOpen
  \bibfield  {author} {\bibinfo {author} {\bibfnamefont {Y.}~\bibnamefont
  {Rosenfeld}},\ }\href {\doibase 10.1103/PhysRevA.15.2545} {\bibfield
  {journal} {\bibinfo  {journal} {Phys. Rev. A}\ }\textbf {\bibinfo {volume}
  {15}},\ \bibinfo {pages} {2545} (\bibinfo {year} {1977})}\BibitemShut
  {NoStop}%
\bibitem [{\citenamefont {Rosenfeld}(1999)}]{Rosenfeld1999}%
  \BibitemOpen
  \bibfield  {author} {\bibinfo {author} {\bibfnamefont {Y.}~\bibnamefont
  {Rosenfeld}},\ }\href {http://stacks.iop.org/0953-8984/11/i=28/a=303}
  {\bibfield  {journal} {\bibinfo  {journal} {J. Phys.: Condens. Matter}\
  }\textbf {\bibinfo {volume} {11}},\ \bibinfo {pages} {5415} (\bibinfo {year}
  {1999})}\BibitemShut {NoStop}%
\bibitem [{\citenamefont {Hoyt}, \citenamefont {Asta},\ and\ \citenamefont
  {Sadigh}(2000)}]{Hoyt2000}%
  \BibitemOpen
  \bibfield  {author} {\bibinfo {author} {\bibfnamefont {J.~J.}\ \bibnamefont
  {Hoyt}}, \bibinfo {author} {\bibfnamefont {M.}~\bibnamefont {Asta}}, \ and\
  \bibinfo {author} {\bibfnamefont {B.}~\bibnamefont {Sadigh}},\ }\href
  {\doibase 10.1103/PhysRevLett.85.594} {\bibfield  {journal} {\bibinfo
  {journal} {Phys. Rev. Lett.}\ }\textbf {\bibinfo {volume} {85}},\ \bibinfo
  {pages} {594} (\bibinfo {year} {2000})}\BibitemShut {NoStop}%
\bibitem [{\citenamefont {Sharma}, \citenamefont {Chakraborty},\ and\
  \citenamefont {Chakravarty}(2006)}]{Sharma06}%
  \BibitemOpen
  \bibfield  {author} {\bibinfo {author} {\bibfnamefont {R.}~\bibnamefont
  {Sharma}}, \bibinfo {author} {\bibfnamefont {S.~N.}\ \bibnamefont
  {Chakraborty}}, \ and\ \bibinfo {author} {\bibfnamefont {C.}~\bibnamefont
  {Chakravarty}},\ }\href {\doibase http://dx.doi.org/10.1063/1.2390710}
  {\bibfield  {journal} {\bibinfo  {journal} {J. Chem. Phys.}\ }\textbf
  {\bibinfo {volume} {125}},\ \bibinfo {eid} {204501} (\bibinfo {year}
  {2006})}\BibitemShut {NoStop}%
\bibitem [{\citenamefont {Mittal}, \citenamefont {Errington},\ and\
  \citenamefont {Truskett}(2006{\natexlab{a}})}]{Mittal2006s2}%
  \BibitemOpen
  \bibfield  {author} {\bibinfo {author} {\bibfnamefont {J.}~\bibnamefont
  {Mittal}}, \bibinfo {author} {\bibfnamefont {J.~R.}\ \bibnamefont
  {Errington}}, \ and\ \bibinfo {author} {\bibfnamefont {T.~M.}\ \bibnamefont
  {Truskett}},\ }\href {\doibase 10.1021/jp064816j} {\bibfield  {journal}
  {\bibinfo  {journal} {J. Phys. Chem. B}\ }\textbf {\bibinfo {volume} {110}},\
  \bibinfo {pages} {18147} (\bibinfo {year} {2006}{\natexlab{a}})}\BibitemShut
  {NoStop}%
\bibitem [{\citenamefont {Mittal}, \citenamefont {Errington},\ and\
  \citenamefont {Truskett}(2006{\natexlab{b}})}]{Mittal2006supercooled}%
  \BibitemOpen
  \bibfield  {author} {\bibinfo {author} {\bibfnamefont {J.}~\bibnamefont
  {Mittal}}, \bibinfo {author} {\bibfnamefont {J.~R.}\ \bibnamefont
  {Errington}}, \ and\ \bibinfo {author} {\bibfnamefont {T.~M.}\ \bibnamefont
  {Truskett}},\ }\href {\doibase http://dx.doi.org/10.1063/1.2336197}
  {\bibfield  {journal} {\bibinfo  {journal} {J. Chem. Phys.}\ }\textbf
  {\bibinfo {volume} {125}},\ \bibinfo {eid} {076102} (\bibinfo {year}
  {2006}{\natexlab{b}})}\BibitemShut {NoStop}%
\bibitem [{\citenamefont {Gnan}\ \emph {et~al.}(2009)\citenamefont {Gnan},
  \citenamefont {Schr\o{}der}, \citenamefont {Pedersen}, \citenamefont
  {Bailey},\ and\ \citenamefont {Dyre}}]{GnanDyre2009}%
  \BibitemOpen
  \bibfield  {author} {\bibinfo {author} {\bibfnamefont {N.}~\bibnamefont
  {Gnan}}, \bibinfo {author} {\bibfnamefont {T.~B.}\ \bibnamefont
  {Schr\o{}der}}, \bibinfo {author} {\bibfnamefont {U.~R.}\ \bibnamefont
  {Pedersen}}, \bibinfo {author} {\bibfnamefont {N.~P.}\ \bibnamefont
  {Bailey}}, \ and\ \bibinfo {author} {\bibfnamefont {J.~C.}\ \bibnamefont
  {Dyre}},\ }\href {\doibase http://dx.doi.org/10.1063/1.3265957} {\bibfield
  {journal} {\bibinfo  {journal} {J. Chem. Phys.}\ }\textbf {\bibinfo {volume}
  {131}},\ \bibinfo {eid} {234504} (\bibinfo {year} {2009})}\BibitemShut
  {NoStop}%
\bibitem [{\citenamefont {Krekelberg}\ \emph {et~al.}(2009)\citenamefont
  {Krekelberg}, \citenamefont {Pond}, \citenamefont {Goel}, \citenamefont
  {Shen}, \citenamefont {Errington},\ and\ \citenamefont
  {Truskett}}]{Krekelberg2009}%
  \BibitemOpen
  \bibfield  {author} {\bibinfo {author} {\bibfnamefont {W.~P.}\ \bibnamefont
  {Krekelberg}}, \bibinfo {author} {\bibfnamefont {M.~J.}\ \bibnamefont
  {Pond}}, \bibinfo {author} {\bibfnamefont {G.}~\bibnamefont {Goel}}, \bibinfo
  {author} {\bibfnamefont {V.~K.}\ \bibnamefont {Shen}}, \bibinfo {author}
  {\bibfnamefont {J.~R.}\ \bibnamefont {Errington}}, \ and\ \bibinfo {author}
  {\bibfnamefont {T.~M.}\ \bibnamefont {Truskett}},\ }\href {\doibase
  10.1103/PhysRevE.80.061205} {\bibfield  {journal} {\bibinfo  {journal} {Phys.
  Rev. E}\ }\textbf {\bibinfo {volume} {80}},\ \bibinfo {pages} {061205}
  (\bibinfo {year} {2009})}\BibitemShut {NoStop}%
\bibitem [{\citenamefont {Pond}, \citenamefont {Errington},\ and\ \citenamefont
  {Truskett}(2011)}]{PondJCP2011}%
  \BibitemOpen
  \bibfield  {author} {\bibinfo {author} {\bibfnamefont {M.~J.}\ \bibnamefont
  {Pond}}, \bibinfo {author} {\bibfnamefont {J.~R.}\ \bibnamefont {Errington}},
  \ and\ \bibinfo {author} {\bibfnamefont {T.~M.}\ \bibnamefont {Truskett}},\
  }\href {\doibase http://dx.doi.org/10.1063/1.3559676} {\bibfield  {journal}
  {\bibinfo  {journal} {J. Chem. Phys.}\ }\textbf {\bibinfo {volume} {134}},\
  \bibinfo {eid} {081101} (\bibinfo {year} {2011})}\BibitemShut {NoStop}%
\bibitem [{\citenamefont {Nayar}\ and\ \citenamefont
  {Chakravarty}(2013)}]{Nayar13}%
  \BibitemOpen
  \bibfield  {author} {\bibinfo {author} {\bibfnamefont {D.}~\bibnamefont
  {Nayar}}\ and\ \bibinfo {author} {\bibfnamefont {C.}~\bibnamefont
  {Chakravarty}},\ }\href {\doibase 10.1039/C3CP51114F} {\bibfield  {journal}
  {\bibinfo  {journal} {Phys. Chem. Chem. Phys.}\ }\textbf {\bibinfo {volume}
  {15}},\ \bibinfo {pages} {14162} (\bibinfo {year} {2013})}\BibitemShut
  {NoStop}%
\bibitem [{\citenamefont {Dyre}(2014)}]{Dyre2014}%
  \BibitemOpen
  \bibfield  {author} {\bibinfo {author} {\bibfnamefont {J.~C.}\ \bibnamefont
  {Dyre}},\ }\href {\doibase 10.1021/jp501852b} {\bibfield  {journal} {\bibinfo
   {journal} {J. Phys. Chem. B}\ }\textbf {\bibinfo {volume} {118}},\ \bibinfo
  {pages} {10007} (\bibinfo {year} {2014})}\BibitemShut {NoStop}%
\bibitem [{\citenamefont {Mittal}, \citenamefont {Errington},\ and\
  \citenamefont {Truskett}(2006{\natexlab{c}})}]{Mittal2006}%
  \BibitemOpen
  \bibfield  {author} {\bibinfo {author} {\bibfnamefont {J.}~\bibnamefont
  {Mittal}}, \bibinfo {author} {\bibfnamefont {J.~R.}\ \bibnamefont
  {Errington}}, \ and\ \bibinfo {author} {\bibfnamefont {T.~M.}\ \bibnamefont
  {Truskett}},\ }\href {\doibase 10.1103/PhysRevLett.96.177804} {\bibfield
  {journal} {\bibinfo  {journal} {Phys. Rev. Lett.}\ }\textbf {\bibinfo
  {volume} {96}},\ \bibinfo {pages} {177804} (\bibinfo {year}
  {2006}{\natexlab{c}})}\BibitemShut {NoStop}%
\bibitem [{\citenamefont {Mittal}, \citenamefont {Errington},\ and\
  \citenamefont {Truskett}(2007)}]{Mittal2007}%
  \BibitemOpen
  \bibfield  {author} {\bibinfo {author} {\bibfnamefont {J.}~\bibnamefont
  {Mittal}}, \bibinfo {author} {\bibfnamefont {J.~R.}\ \bibnamefont
  {Errington}}, \ and\ \bibinfo {author} {\bibfnamefont {T.~M.}\ \bibnamefont
  {Truskett}},\ }\href {\doibase 10.1021/jp071369e} {\bibfield  {journal}
  {\bibinfo  {journal} {J. Phys. Chem. B}\ }\textbf {\bibinfo {volume} {111}},\
  \bibinfo {pages} {10054} (\bibinfo {year} {2007})}\BibitemShut {NoStop}%
\bibitem [{\citenamefont {Mittal}\ \emph {et~al.}(2007)\citenamefont {Mittal},
  \citenamefont {Shen}, \citenamefont {Errington},\ and\ \citenamefont
  {Truskett}}]{Mittal2007mixtures}%
  \BibitemOpen
  \bibfield  {author} {\bibinfo {author} {\bibfnamefont {J.}~\bibnamefont
  {Mittal}}, \bibinfo {author} {\bibfnamefont {V.~K.}\ \bibnamefont {Shen}},
  \bibinfo {author} {\bibfnamefont {J.~R.}\ \bibnamefont {Errington}}, \ and\
  \bibinfo {author} {\bibfnamefont {T.~M.}\ \bibnamefont {Truskett}},\ }\href
  {\doibase http://dx.doi.org/10.1063/1.2795699} {\bibfield  {journal}
  {\bibinfo  {journal} {J. Chem. Phys.}\ }\textbf {\bibinfo {volume} {127}},\
  \bibinfo {eid} {154513} (\bibinfo {year} {2007})}\BibitemShut {NoStop}%
\bibitem [{\citenamefont {Goel}\ \emph {et~al.}(2008)\citenamefont {Goel},
  \citenamefont {Krekelberg}, \citenamefont {Errington},\ and\ \citenamefont
  {Truskett}}]{Goel2008}%
  \BibitemOpen
  \bibfield  {author} {\bibinfo {author} {\bibfnamefont {G.}~\bibnamefont
  {Goel}}, \bibinfo {author} {\bibfnamefont {W.~P.}\ \bibnamefont
  {Krekelberg}}, \bibinfo {author} {\bibfnamefont {J.~R.}\ \bibnamefont
  {Errington}}, \ and\ \bibinfo {author} {\bibfnamefont {T.~M.}\ \bibnamefont
  {Truskett}},\ }\href {\doibase 10.1103/PhysRevLett.100.106001} {\bibfield
  {journal} {\bibinfo  {journal} {Phys. Rev. Lett.}\ }\textbf {\bibinfo
  {volume} {100}},\ \bibinfo {pages} {106001} (\bibinfo {year}
  {2008})}\BibitemShut {NoStop}%
\bibitem [{\citenamefont {Goel}\ \emph {et~al.}(2009)\citenamefont {Goel},
  \citenamefont {Krekelberg}, \citenamefont {Pond}, \citenamefont {Mittal},
  \citenamefont {Shen}, \citenamefont {Errington},\ and\ \citenamefont
  {Truskett}}]{Goel2009}%
  \BibitemOpen
  \bibfield  {author} {\bibinfo {author} {\bibfnamefont {G.}~\bibnamefont
  {Goel}}, \bibinfo {author} {\bibfnamefont {W.~P.}\ \bibnamefont
  {Krekelberg}}, \bibinfo {author} {\bibfnamefont {M.~J.}\ \bibnamefont
  {Pond}}, \bibinfo {author} {\bibfnamefont {J.}~\bibnamefont {Mittal}},
  \bibinfo {author} {\bibfnamefont {V.~K.}\ \bibnamefont {Shen}}, \bibinfo
  {author} {\bibfnamefont {J.~R.}\ \bibnamefont {Errington}}, \ and\ \bibinfo
  {author} {\bibfnamefont {T.~M.}\ \bibnamefont {Truskett}},\ }\href
  {http://stacks.iop.org/1742-5468/2009/i=04/a=P04006} {\bibfield  {journal}
  {\bibinfo  {journal} {J. Stat. Mech.: Theory Exp.}\ }\textbf {\bibinfo
  {volume} {2009}},\ \bibinfo {pages} {P04006} (\bibinfo {year}
  {2009})}\BibitemShut {NoStop}%
\bibitem [{\citenamefont {Chopra}, \citenamefont {Truskett},\ and\
  \citenamefont {Errington}(2010)}]{Chopra2010confined}%
  \BibitemOpen
  \bibfield  {author} {\bibinfo {author} {\bibfnamefont {R.}~\bibnamefont
  {Chopra}}, \bibinfo {author} {\bibfnamefont {T.~M.}\ \bibnamefont
  {Truskett}}, \ and\ \bibinfo {author} {\bibfnamefont {J.~R.}\ \bibnamefont
  {Errington}},\ }\href {\doibase 10.1103/PhysRevE.82.041201} {\bibfield
  {journal} {\bibinfo  {journal} {Phys. Rev. E}\ }\textbf {\bibinfo {volume}
  {82}},\ \bibinfo {pages} {041201} (\bibinfo {year} {2010})}\BibitemShut
  {NoStop}%
\bibitem [{\citenamefont {Borah}\ \emph {et~al.}(2012)\citenamefont {Borah},
  \citenamefont {Maiti}, \citenamefont {Chakravarty},\ and\ \citenamefont
  {Yashonath}}]{Borah12}%
  \BibitemOpen
  \bibfield  {author} {\bibinfo {author} {\bibfnamefont {B.~J.}\ \bibnamefont
  {Borah}}, \bibinfo {author} {\bibfnamefont {P.~K.}\ \bibnamefont {Maiti}},
  \bibinfo {author} {\bibfnamefont {C.}~\bibnamefont {Chakravarty}}, \ and\
  \bibinfo {author} {\bibfnamefont {S.}~\bibnamefont {Yashonath}},\ }\href
  {\doibase http://dx.doi.org/10.1063/1.4706520} {\bibfield  {journal}
  {\bibinfo  {journal} {J. Chem. Phys.}\ }\textbf {\bibinfo {volume} {136}},\
  \bibinfo {eid} {174510} (\bibinfo {year} {2012})}\BibitemShut {NoStop}%
\bibitem [{\citenamefont {Ma}\ \emph {et~al.}(2013)\citenamefont {Ma},
  \citenamefont {Chen}, \citenamefont {Wang}, \citenamefont {Peng},
  \citenamefont {Han},\ and\ \citenamefont {Tong}}]{Ma2013}%
  \BibitemOpen
  \bibfield  {author} {\bibinfo {author} {\bibfnamefont {X.}~\bibnamefont
  {Ma}}, \bibinfo {author} {\bibfnamefont {W.}~\bibnamefont {Chen}}, \bibinfo
  {author} {\bibfnamefont {Z.}~\bibnamefont {Wang}}, \bibinfo {author}
  {\bibfnamefont {Y.}~\bibnamefont {Peng}}, \bibinfo {author} {\bibfnamefont
  {Y.}~\bibnamefont {Han}}, \ and\ \bibinfo {author} {\bibfnamefont
  {P.}~\bibnamefont {Tong}},\ }\href {\doibase 10.1103/PhysRevLett.110.078302}
  {\bibfield  {journal} {\bibinfo  {journal} {Phys. Rev. Lett.}\ }\textbf
  {\bibinfo {volume} {110}},\ \bibinfo {pages} {078302} (\bibinfo {year}
  {2013})}\BibitemShut {NoStop}%
\bibitem [{\citenamefont {Liu}, \citenamefont {Fu},\ and\ \citenamefont
  {Wu}(2013)}]{Liu2013}%
  \BibitemOpen
  \bibfield  {author} {\bibinfo {author} {\bibfnamefont {Y.}~\bibnamefont
  {Liu}}, \bibinfo {author} {\bibfnamefont {J.}~\bibnamefont {Fu}}, \ and\
  \bibinfo {author} {\bibfnamefont {J.}~\bibnamefont {Wu}},\ }\href {\doibase
  10.1021/la403082q} {\bibfield  {journal} {\bibinfo  {journal} {Langmuir}\
  }\textbf {\bibinfo {volume} {29}},\ \bibinfo {pages} {12997} (\bibinfo {year}
  {2013})}\BibitemShut {NoStop}%
\bibitem [{\citenamefont {Ingebrigtsen}\ \emph {et~al.}(2013)\citenamefont
  {Ingebrigtsen}, \citenamefont {Errington}, \citenamefont {Truskett},\ and\
  \citenamefont {Dyre}}]{Ingebrigtsen2013}%
  \BibitemOpen
  \bibfield  {author} {\bibinfo {author} {\bibfnamefont {T.~S.}\ \bibnamefont
  {Ingebrigtsen}}, \bibinfo {author} {\bibfnamefont {J.~R.}\ \bibnamefont
  {Errington}}, \bibinfo {author} {\bibfnamefont {T.~M.}\ \bibnamefont
  {Truskett}}, \ and\ \bibinfo {author} {\bibfnamefont {J.~C.}\ \bibnamefont
  {Dyre}},\ }\href {\doibase 10.1103/PhysRevLett.111.235901} {\bibfield
  {journal} {\bibinfo  {journal} {Phys. Rev. Lett.}\ }\textbf {\bibinfo
  {volume} {111}},\ \bibinfo {pages} {235901} (\bibinfo {year}
  {2013})}\BibitemShut {NoStop}%
\bibitem [{\citenamefont {Edmond}, \citenamefont {Nugent},\ and\ \citenamefont
  {Weeks}(2012)}]{Edmond2012}%
  \BibitemOpen
  \bibfield  {author} {\bibinfo {author} {\bibfnamefont {K.~V.}\ \bibnamefont
  {Edmond}}, \bibinfo {author} {\bibfnamefont {C.~R.}\ \bibnamefont {Nugent}},
  \ and\ \bibinfo {author} {\bibfnamefont {E.~R.}\ \bibnamefont {Weeks}},\
  }\href {\doibase 10.1103/PhysRevE.85.041401} {\bibfield  {journal} {\bibinfo
  {journal} {Phys. Rev. E}\ }\textbf {\bibinfo {volume} {85}},\ \bibinfo
  {pages} {041401} (\bibinfo {year} {2012})}\BibitemShut {NoStop}%
\bibitem [{\citenamefont {Bollinger}, \citenamefont {Jain},\ and\ \citenamefont
  {Truskett}(2014{\natexlab{a}})}]{Bollinger2014}%
  \BibitemOpen
  \bibfield  {author} {\bibinfo {author} {\bibfnamefont {J.~A.}\ \bibnamefont
  {Bollinger}}, \bibinfo {author} {\bibfnamefont {A.}~\bibnamefont {Jain}}, \
  and\ \bibinfo {author} {\bibfnamefont {T.~M.}\ \bibnamefont {Truskett}},\
  }\href {\doibase 10.1021/la5017005} {\bibfield  {journal} {\bibinfo
  {journal} {Langmuir}\ }\textbf {\bibinfo {volume} {30}},\ \bibinfo {pages}
  {8247} (\bibinfo {year} {2014}{\natexlab{a}})}\BibitemShut {NoStop}%
\bibitem [{\citenamefont {Bollinger}, \citenamefont {Jain},\ and\ \citenamefont
  {Truskett}(2014{\natexlab{b}})}]{BollingerJPCB2014}%
  \BibitemOpen
  \bibfield  {author} {\bibinfo {author} {\bibfnamefont {J.~A.}\ \bibnamefont
  {Bollinger}}, \bibinfo {author} {\bibfnamefont {A.}~\bibnamefont {Jain}}, \
  and\ \bibinfo {author} {\bibfnamefont {T.~M.}\ \bibnamefont {Truskett}},\
  }\href {\doibase DOI: 10.1021/jp508887r} {\bibfield  {journal} {\bibinfo
  {journal} {J. Phys. Chem. B}\ } (\bibinfo {year} {2014}{\natexlab{b}}),\ DOI:
  10.1021/jp508887r}\BibitemShut {NoStop}%
\bibitem [{\citenamefont {Chandler}, \citenamefont {Weeks},\ and\ \citenamefont
  {Andersen}(1983)}]{ChandlerWeeksAndersenScience1983}%
  \BibitemOpen
  \bibfield  {author} {\bibinfo {author} {\bibfnamefont {D.}~\bibnamefont
  {Chandler}}, \bibinfo {author} {\bibfnamefont {J.~D.}\ \bibnamefont {Weeks}},
  \ and\ \bibinfo {author} {\bibfnamefont {H.~C.}\ \bibnamefont {Andersen}},\
  }\href {\doibase 10.1126/science.220.4599.787} {\ \textbf {\bibinfo {volume}
  {220}},\ \bibinfo {pages} {787} (\bibinfo {year} {1983})}\BibitemShut
  {NoStop}%
\bibitem [{\citenamefont {Hess}\ \emph {et~al.}(2008)\citenamefont {Hess},
  \citenamefont {Kutzner}, \citenamefont {van~der Spoel},\ and\ \citenamefont
  {Lindahl}}]{HessJCTC2008}%
  \BibitemOpen
  \bibfield  {author} {\bibinfo {author} {\bibfnamefont {B.}~\bibnamefont
  {Hess}}, \bibinfo {author} {\bibfnamefont {C.}~\bibnamefont {Kutzner}},
  \bibinfo {author} {\bibfnamefont {D.}~\bibnamefont {van~der Spoel}}, \ and\
  \bibinfo {author} {\bibfnamefont {E.}~\bibnamefont {Lindahl}},\ }\href
  {\doibase 10.1021/ct700301q} {\bibfield  {journal} {\bibinfo  {journal} {J.
  Chem. Theory Comput.}\ }\textbf {\bibinfo {volume} {4}},\ \bibinfo {pages}
  {435} (\bibinfo {year} {2008})}\BibitemShut {NoStop}%
\bibitem [{Note1()}]{Note1}%
  \BibitemOpen
  \bibinfo {note} {See supplemental material at [URL will be inserted by AIP]
  for detailed simulation protocols and auxiliary data.}\BibitemShut {Stop}%
\bibitem [{\citenamefont {Liu}, \citenamefont {Harder},\ and\ \citenamefont
  {Berne}(2004)}]{LiuBerne2004}%
  \BibitemOpen
  \bibfield  {author} {\bibinfo {author} {\bibfnamefont {P.}~\bibnamefont
  {Liu}}, \bibinfo {author} {\bibfnamefont {E.}~\bibnamefont {Harder}}, \ and\
  \bibinfo {author} {\bibfnamefont {B.~J.}\ \bibnamefont {Berne}},\ }\href
  {\doibase 10.1021/jp0375057} {\bibfield  {journal} {\bibinfo  {journal} {J.
  Phys. Chem. B}\ }\textbf {\bibinfo {volume} {108}},\ \bibinfo {pages} {6595}
  (\bibinfo {year} {2004})}\BibitemShut {NoStop}%
\bibitem [{\citenamefont {Weiss}(2007)}]{WeissACP1966}%
  \BibitemOpen
  \bibfield  {author} {\bibinfo {author} {\bibfnamefont {G.~H.}\ \bibnamefont
  {Weiss}},\ }\enquote {\bibinfo {title} {First passage time problems in
  chemical physics},}\ in\ \href {\doibase 10.1002/9780470140154.ch1} {\emph
  {\bibinfo {booktitle} {Adv. Chem. Phys.}}}\ (\bibinfo  {publisher} {John
  Wiley \& Sons, Inc.},\ \bibinfo {year} {2007})\ pp.\ \bibinfo {pages}
  {1--18}\BibitemShut {NoStop}%
\bibitem [{\citenamefont {Sedlmeier}\ \emph {et~al.}(2011)\citenamefont
  {Sedlmeier}, \citenamefont {Hansen}, \citenamefont {Mengyu}, \citenamefont
  {Horinek},\ and\ \citenamefont {Netz}}]{Netz2011}%
  \BibitemOpen
  \bibfield  {author} {\bibinfo {author} {\bibfnamefont {F.}~\bibnamefont
  {Sedlmeier}}, \bibinfo {author} {\bibfnamefont {Y.}~\bibnamefont {Hansen}},
  \bibinfo {author} {\bibfnamefont {L.}~\bibnamefont {Mengyu}}, \bibinfo
  {author} {\bibfnamefont {D.}~\bibnamefont {Horinek}}, \ and\ \bibinfo
  {author} {\bibfnamefont {R.}~\bibnamefont {Netz}},\ }\href {\doibase
  10.1007/s10955-011-0338-0} {\bibfield  {journal} {\bibinfo  {journal} {J.
  Stat. Phys.}\ }\textbf {\bibinfo {volume} {145}},\ \bibinfo {pages} {240}
  (\bibinfo {year} {2011})}\BibitemShut {NoStop}%
\bibitem [{\citenamefont {Widom}(1963)}]{Widom1963}%
  \BibitemOpen
  \bibfield  {author} {\bibinfo {author} {\bibfnamefont {B.}~\bibnamefont
  {Widom}},\ }\href@noop {} {\bibfield  {journal} {\bibinfo  {journal} {J.
  Chem. Phys.}\ }\textbf {\bibinfo {volume} {39}} (\bibinfo {year}
  {1963})}\BibitemShut {NoStop}%
\bibitem [{\citenamefont {Sastry}\ \emph {et~al.}(1998)\citenamefont {Sastry},
  \citenamefont {Truskett}, \citenamefont {Debenedetti}, \citenamefont
  {Torquato},\ and\ \citenamefont {Stillinger}}]{Sastry1998}%
  \BibitemOpen
  \bibfield  {author} {\bibinfo {author} {\bibfnamefont {S.}~\bibnamefont
  {Sastry}}, \bibinfo {author} {\bibfnamefont {T.~M.}\ \bibnamefont
  {Truskett}}, \bibinfo {author} {\bibfnamefont {P.~G.}\ \bibnamefont
  {Debenedetti}}, \bibinfo {author} {\bibfnamefont {S.}~\bibnamefont
  {Torquato}}, \ and\ \bibinfo {author} {\bibfnamefont {F.~H.}\ \bibnamefont
  {Stillinger}},\ }\href {\doibase 10.1080/00268979809483161} {\bibfield
  {journal} {\bibinfo  {journal} {Mol. Phys.}\ }\textbf {\bibinfo {volume}
  {95}},\ \bibinfo {pages} {289} (\bibinfo {year} {1998})}\BibitemShut
  {NoStop}%
\bibitem [{\citenamefont {Weeks}\ and\ \citenamefont
  {Weitz}(2002)}]{WeeksWeitz2002}%
  \BibitemOpen
  \bibfield  {author} {\bibinfo {author} {\bibfnamefont {E.~R.}\ \bibnamefont
  {Weeks}}\ and\ \bibinfo {author} {\bibfnamefont {D.~A.}\ \bibnamefont
  {Weitz}},\ }\href {\doibase 10.1103/PhysRevLett.89.095704} {\bibfield
  {journal} {\bibinfo  {journal} {Phys. Rev. Lett.}\ }\textbf {\bibinfo
  {volume} {89}},\ \bibinfo {pages} {095704} (\bibinfo {year}
  {2002})}\BibitemShut {NoStop}%
\bibitem [{\citenamefont {Ermak}\ and\ \citenamefont
  {McCammon}(1978)}]{Ermak1978}%
  \BibitemOpen
  \bibfield  {author} {\bibinfo {author} {\bibfnamefont {D.~L.}\ \bibnamefont
  {Ermak}}\ and\ \bibinfo {author} {\bibfnamefont {J.~A.}\ \bibnamefont
  {McCammon}},\ }\href {\doibase http://dx.doi.org/10.1063/1.436761} {\bibfield
   {journal} {\bibinfo  {journal} {J. Chem. Phys.}\ }\textbf {\bibinfo {volume}
  {69}},\ \bibinfo {pages} {1352} (\bibinfo {year} {1978})}\BibitemShut
  {NoStop}%
\bibitem [{\citenamefont {Allen}\ and\ \citenamefont
  {Tildesley}(1989)}]{AllenTildesley1989}%
  \BibitemOpen
  \bibfield  {author} {\bibinfo {author} {\bibfnamefont {M.~P.}\ \bibnamefont
  {Allen}}\ and\ \bibinfo {author} {\bibfnamefont {D.~J.}\ \bibnamefont
  {Tildesley}},\ }\href@noop {} {\emph {\bibinfo {title} {Computer Simulation
  of Liquids}}}\ (\bibinfo  {publisher} {Clarendon Press},\ \bibinfo {address}
  {New York, NY, USA},\ \bibinfo {year} {1989})\BibitemShut {NoStop}%
\bibitem [{\citenamefont {Lubachevsky}\ and\ \citenamefont
  {Stillinger}(1990)}]{LubachevskyStillinger1990}%
  \BibitemOpen
  \bibfield  {author} {\bibinfo {author} {\bibfnamefont {B.~D.}\ \bibnamefont
  {Lubachevsky}}\ and\ \bibinfo {author} {\bibfnamefont {F.~H.}\ \bibnamefont
  {Stillinger}},\ }\href {\doibase 10.1007/BF01025983} {\bibfield  {journal}
  {\bibinfo  {journal} {J. Stat. Phys.}\ }\textbf {\bibinfo {volume} {60}},\
  \bibinfo {pages} {561} (\bibinfo {year} {1990})}\BibitemShut {NoStop}%
\bibitem [{\citenamefont {Hansen}\ and\ \citenamefont
  {McDonald}(2006)}]{HansenMcDonald2006}%
  \BibitemOpen
  \bibfield  {author} {\bibinfo {author} {\bibfnamefont {J.-P.}\ \bibnamefont
  {Hansen}}\ and\ \bibinfo {author} {\bibfnamefont {I.~R.}\ \bibnamefont
  {McDonald}},\ }\href@noop {} {\emph {\bibinfo {title} {Theory of Simple
  Liquids}}},\ \bibinfo {edition} {3rd}\ ed.\ (\bibinfo  {publisher} {Academic
  Press},\ \bibinfo {address} {New York, NY, USA},\ \bibinfo {year}
  {2006})\BibitemShut {NoStop}%
\bibitem [{\citenamefont {Shen}\ and\ \citenamefont
  {Errington}(2005)}]{Shen2005}%
  \BibitemOpen
  \bibfield  {author} {\bibinfo {author} {\bibfnamefont {V.~K.}\ \bibnamefont
  {Shen}}\ and\ \bibinfo {author} {\bibfnamefont {J.~R.}\ \bibnamefont
  {Errington}},\ }\href {\doibase http://dx.doi.org/10.1063/1.1844372}
  {\bibfield  {journal} {\bibinfo  {journal} {J. Chem. Phys.}\ }\textbf
  {\bibinfo {volume} {122}},\ \bibinfo {eid} {064508} (\bibinfo {year}
  {2005})}\BibitemShut {NoStop}%

\end{thebibliography}
\end{document}